\documentclass[twocolumn,showpacs,preprintnumbers,amsmath,amssymb]{revtex4}
\usepackage{epsfig}
\usepackage{graphicx}
\usepackage{dcolumn}
\usepackage{bm}
\usepackage{float}

\begin{document}

\selectfont

\title{Inter-arrival times of message propagation on directed networks}

\author{Tamara Mihaljev}
\email{tamaram@ethz.ch}
\affiliation{Computational Physics, IfB, ETH Zurich, Schafmattstrasse 6, 8093 Zurich, Switzerland}

\author{Lucilla de Arcangelis}
\email{dearcangelis@na.infn.it}
\affiliation{Department of Information Engineering and CNISM, Second University of Naples, 81031 Aversa (CE), Italy}

\author{Hans J. Herrmann}
\affiliation{Computational Physics, IfB, ETH Zurich, Schafmattstrasse 6, 8093 Zurich, Switzerland}
\affiliation{Departamento de F\'{\i}sica, Universidade Federal do Cear\'a, 60451-970 Fortaleza, Cear\'a, Brazil}

\date{\today}
 
\begin{abstract}
One of the challenges in fighting cybercrime is to understand the dynamics of message 
propagation on botnets, networks of infected computers used to send viruses, unsolicited 
commercial emails (SPAM) or denial of service attacks. We map this problem to the propagation 
of multiple random walkers on directed networks and we evaluate the inter-arrival time 
distribution between successive walkers arriving at a target. We show that the temporal 
organization of this process, which models information propagation on unstructured peer to 
peer networks, has the same features as SPAM arriving to a single user. We study the behavior 
of the message inter-arrival time distribution on three different network topologies using 
two different rules for sending messages. In all networks the propagation is not a pure 
Poisson process. It shows universal features on Poissonian networks and a more complex 
behavior on scale free networks. Results open the possibility to indirectly learn about the 
process of sending messages on networks with unknown topologies, by studying inter-arrival 
times at any node of the network.  
\end{abstract}

 

\maketitle


\section{Introduction}

A botnet is a network of infected computers which are under command and control of a single 
person, the "botmaster".  Botnets are abstract overlay networks on top of the physical 
network topology. They are used for sending unsolicited commercial emails (SPAM), viruses, 
denial of service attacks, for stealing identity data, and for other sorts of cyber crime.  
Botnets are the primary security threat on Internet today \cite{eu}. They grew up to be a 
global and 
multi million dollar business. Fighting botnets is a hard task since their structure is 
constantly evolving, and their inner working is not known. Understanding the dynamics of 
communication on such networks is a big challenge and could be crucial for finding effective 
tools to fighting botnets. This problem can be successfully approached from the network theory 
perspective.

Botnets can have different structures. The new generations of botnets, which are more robust 
against attacks and very difficult to track, are based on peer to peer communication 
\cite{p2pBotnets}. This type of communication can be modeled as a random walk. When the 
botmaster sends an order to its bots, it sends it only to a fraction of nodes in the botnet, 
which can then forward it to only those bots whose IP addresses they know. These addresses 
are randomly assigned to bots. There is no centralized point in such networks since all the 
nodes are equally important, they are all clients and servers at the same time. This is the 
reason for botnet's robustness against attacks. Even if a node of the botnet is identified, 
its communication with the rest of the botnet can be tracked back only to a limited number of 
bots. Therefore, no attack would destroy the whole botnet, or lead to the botmaster. Links in 
peer to peer networks do not have to be bidirectional, and the unstructured peer to peer 
networks, often used for constructing botnets since their are the most difficult to track, 
have random topology. Therefore we will model botnets as random directed networks, and we 
will study the propagaton of random walkers on them in order to attack the problem of 
understanding the internal mechanisms of message propagation on these networks.
Random walks on directed networks are also an interesting fundamental problem, important 
for understanding the communication in any other peer to peer network \cite{otherP2P}, 
wireless  sensor networks \cite{directedWirelessRouting}, ad hoc networks \cite{adHocNets}, 
or different processes on world wide web, such as tagging \cite{loretoTagging}. The results 
we present are general, not botnet specific, and are valid for any system in which data 
packets propagate  in a random fashion on the directed network.

Random walks and related stochastic processes have mainly been studied on regular lattices 
and $d$-dimensional euclidian spaces in the past \cite{rwBook}, due to their obvious 
relevance to physical problems. In recent years networks are becoming the preferred model to 
study complex systems \cite{boccaletti, vespignani} and this triggered studies of random 
walks on them \cite{riegerRWonComplexNets,gallosRandomTrapping,klafterNature,rednerFPTerdosRenyi, shlomoBook,ladaAdamic,kim,sneppen,germano,bosaTransport,jaspersenBlumen,newmanRWbetweeness,zhouRwCommDet1,zhouRwCommDet2,vishalRWbiased}. 
However, most of the previous results concern random walks on undirected networks. Random 
walks on directed networks have been mainly investigated to find communities in citation 
networks \cite{jeongDirectedCitationNets}, identify subgraph structures on World Wide 
Web \cite{bosaDirected}, or in calculations of the PageRank 
\cite{fortunatoPagerank,caldarelli}. 
This is a measure used by the search engine Google (as well as by several other search engines)
 to determine the prestige of Web pages. When a user submits a query, the hits returned by 
Google are ranked according to the value of their PageRank. The algorithm determining 
this value is based on a modified random walk on the web graph, were nodes are web pages and 
edges between them are naturally directed hyperlinks connecting web pages. In each step the 
modified random walker either follows a randomly chosen outgoing link of the present node, or 
with a small probability, called the damping factor, it jumps to a randomly chosen node in 
the network.

We study random walks on directed networks to model peer to peer communication on botnets 
as spreading of messages. We are interested in the temporal organization of this dynamic 
process and therefore we investigate the distribution of inter-arrival times between two 
successive messages arriving to a given receiver. The inter-arrival time distribution has 
been first introduced to characterize the temporal occurrence of earthquakes 
\cite{BakEarthquakes}. It has been then studied in the contexts of different stochastic 
processes, as solar flares \cite{solarFlares,lucilla2,lucilla3,lucilla4}, 
forest fires\cite{forestFires} or in 
package transport in computer science \cite{inter-arr1, inter-arr2}. The interesting property 
of this quantity is that it is able to provide information about the temporal organization of 
processes whose detailed mechanisms are unknown. In particular, it is a simple exponential if 
the process is Poissonian and therefore it is able to enlighten the presence of temporal 
correlations among events, when a non-exponential behavior is detected.

A recent paper \cite{lucillaSpam}, has investigated the statistical properties of the 
SPAM delivery inter-arrival times. Results have suggested that SPAM messages delivered to a given recipient are time correlated: if the inter-arrival time between two consecutive SPAM messages is small (large), then the next SPAM message will most probably arrive after a small (large) inter-arrival time. SPAM temporal correlations have been reproduced by a numerical model based on the random superposition of SPAM sequences, each one described by the Omori law \cite{omoriLaw}. This and other experimental findings \cite{lucillaSpam} suggest that statistical approaches may be used to infer how spammers operate.

Our motivation to study the distribution of message inter-arrival times on model networks is to detect the eventual presence of temporal correlations and their relation with the network topology. The inter-arrival time distribution of messages sent only to one or a small fraction of nodes could then provide information about the dynamical process taking place on a real network. In the case of botnets this would imply that we would be able to get information about their organization and structure by studying inter-arrival times of either SPAM emails or of contaminated packages, both sent by a botnet, by analyzing data even of a single user. These data are easily accessible, cheap and easy to monitor. Since botnets are difficult to identify, this indirect way of learning about their organization would give a boost in fighting botnets and cyber crime in general.

The paper is organized as follows: In the next section we will describe the model and the implemented networks. In the third section we discuss results obtained for random networks and show their comparison with the real data. In the following two sections we extend our investigation to networks with random topologies without dead ends and to scale free networks.  Finally in the last section we discuss the results and give some concluding remarks.

\section{Model}
\label{2}

We start by constructing a randomly connected directed network with a given degree 
distribution of inputs and outputs. We implement the Poisson distribution, the Poisson 
without dead ends and the power law distributions. We choose a random node in the network to 
be the one where we measure the inter-arrival times between two successive random walker 
arrivals,  $dt=t_{i+1}-t_{i}$. We call this node the target node or the receiver. The target 
node is chosen at random among nodes with a given number of inputs. The number of outputs of the target is not fixed since it does not influence the number of message arrivals. Next, we choose a node from which messages depart, the sender. This node is chosen at random, with a fixed number of outputs. Since only the number of outputs determines the number of different ways a message can leave on its way to the receiver, we do not fix the number of inputs of a sender. In our process the sender is the botmaster sending orders to its bots, and what we measure at the receiver can be compared to the arrivals of messages to a generic user. We have found that increasing the number of senders does not affect the basic properties of the inter-arrival time distribution of messages reaching the receiver.

After a message departs from the sender, it follows at each time step one randomly chosen 
outgoing link of the occupied node. The walk continues until either the target node is reached, or an initially fixed maximal number of steps is exceeded. It is necessary to introduce a limit on the number of steps since, due to the fact that the links are directed, the network may have regions that the walker can enter but cannot escape from. This limit also exists in real internet protocols where it is called the \emph{time to live} (TTL).
This is a limit on the number of transmissions that a data package can experience before it should be discarded. In all simulations presented here we fix this value to twice the size of the network.

To simulate the dynamical process typical for a botmaster, who sends a large number of messages through botnets, many random walkers depart from the sender. We send them either one by one or all at once. The walkers are independent. We record the times at which messages arrive to the receiver and calculate the inter-arrival times between successive messages. When the messages are sent all at once, their arrival times depend solely on the length of the paths undertaken on their way to the receiver. This is mainly affected by the network's topology. When the messages are sent one after another, each consecutive message has an equal time delay in starting its walk to the target node. This process is more complex than the previous one since a message can arrive to the target before others sent earlier due to a shorter undertaken path. In the beginning of the process the number of messages arriving to the target increases with time, until a stationary state is reached. We are interested in the stationary state of the process, where we measure the distribution of inter-arrival times. In our model at each time step a new walker departs from the sender. We have, however, verified that the introduction of a longer time delay between successive departures does not affect the main properties of the distribution.

\section{Random networks} \label{3}

All the networks randomly connect input and output links assigned to the nodes according to some distribution \cite{configModel,newmanCM}. We call, however, the networks random only when the distribution is Poissonian, $p(k)= (\langle k \rangle^k/k!)\cdot e^{\langle k \rangle}$, where $k$ is the number of links and $\langle k \rangle$ its mean value \cite{Bollobas}.
When we construct random networks we choose both, the distribution of input and of output links, to be Poissonian with the same mean degree. The number of walkers has to be large enough to provide good statistics for the distribution of inter-arrival times. 

We sample data from 500 different network realizations for a given degree distribution and  fixed values of the sender outdegree and the receiver indegree. We fix both these values equal to 4. We have, however, verified that the specific value of these parameters does not affect the behavior of the distribution. The space of possible topologies and possible choices of a sender-receiver couple is extremely large. To get better statistics we also fix the distance between these two nodes, namely the shortest path between them.

  \begin{figure}
 \includegraphics*[width=0.45\textwidth]{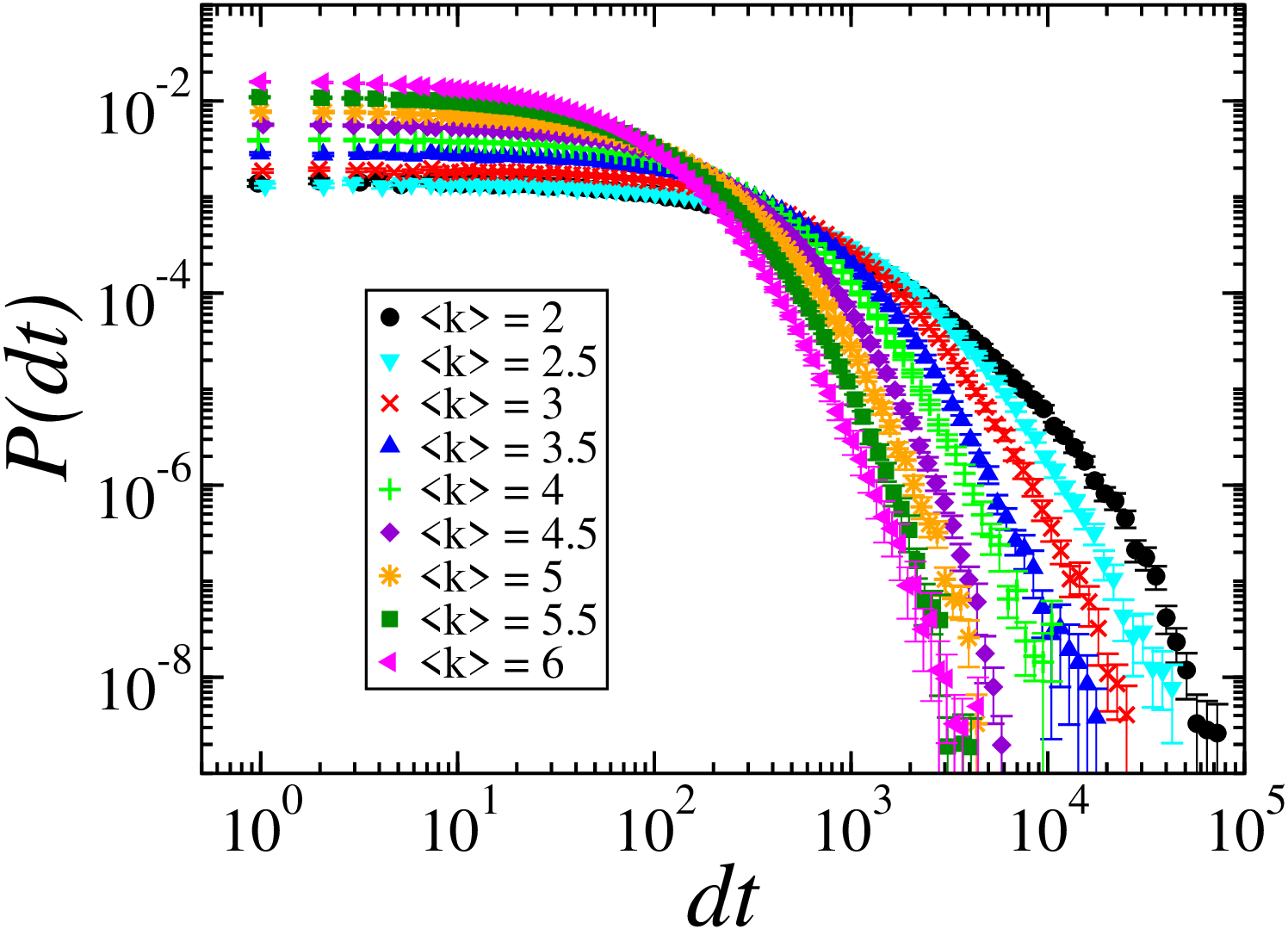}
 \caption{
 Distribution of inter-arrival times in networks with Poisson degree distribution and 
different values of the mean degree. The networks have $N=10^4$ nodes, the number of sent 
messages is $M=10^6$ and the distance between the sender and the receiver is fixed to 8. 
The messages are sent one by one.
 }
 \label{poissDiffK}
 \end{figure}

\begin{figure}[t!]
\includegraphics*[width=0.45\textwidth]{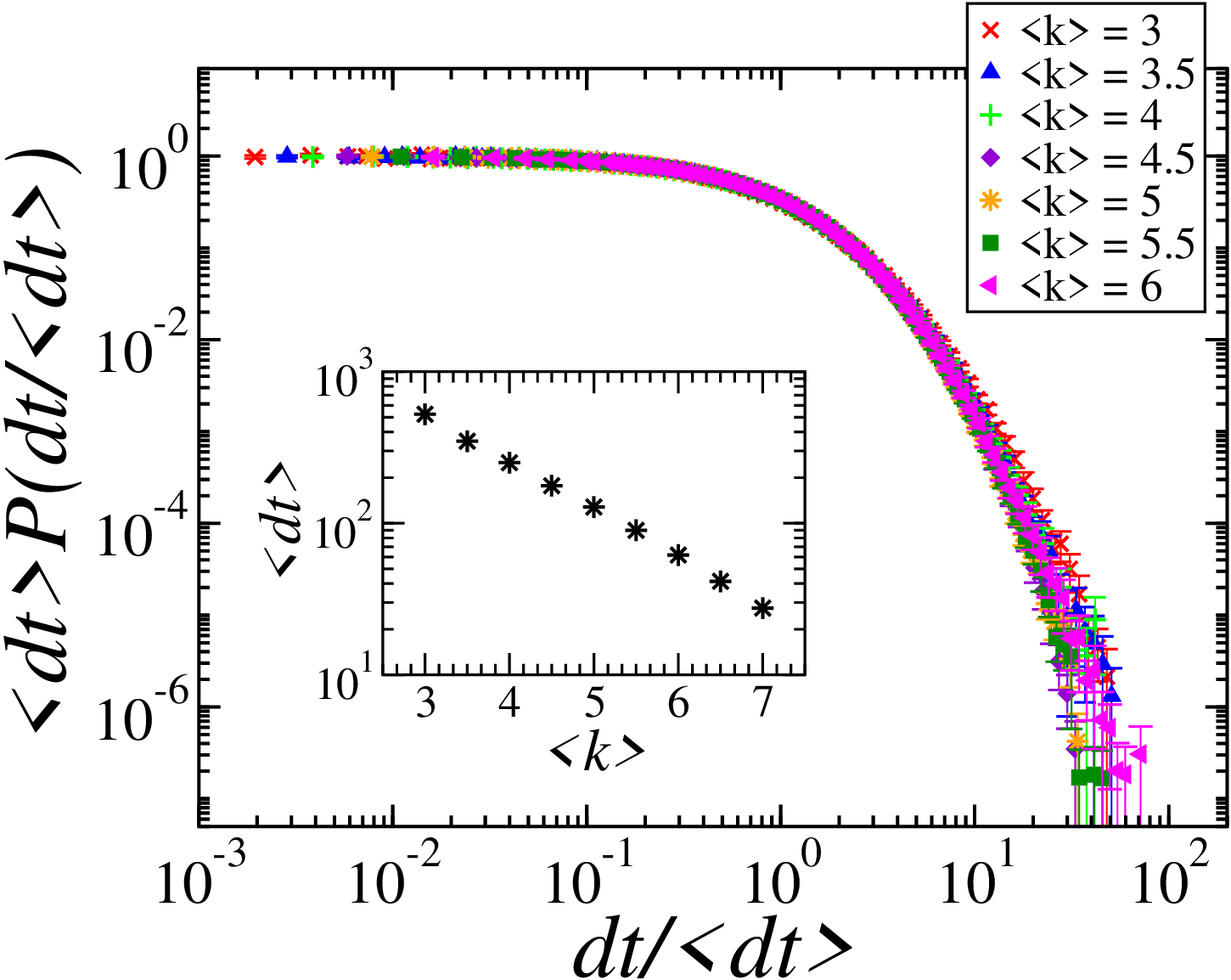}
\caption{
Distributions of inter-arrival times in networks with Poisson degree distribution and different values of the mean degree, rescaled by the average rate of message arrivals. The networks have $N=10^4$ nodes, the number of sent messages is $M=10^6$ and the distance between the sender and the receiver is fixed equal to 8. The messages are sent one by one. The inset shows the dependence of the mean inter-arrival time on the average degree of the network.
}
\label{poissDiffKscaled}
\end{figure}


We first study the case when messages are sent one by one. We find that the distributions of 
inter-arrival times for networks with Poisson distributed links and different average degree 
$\langle k \rangle$ exhibit similar behavior (Fig.\ref{poissDiffK}). Only for networks with 
a small value of the average degree, and therefore a lower level of connectivity, longer 
inter-arrival times are measured. The average inter-arrival time indeed increases 
exponentially as the average connectivity, $\langle k \rangle$, decreases (inset 
Fig.\ref{poissDiffKscaled}). 
In order to check if the distribution is a universal function, solely controlled by the 
average rate $R$ of walkers arriving at the target, we verify the 
following scaling relation \cite{corralInterarrival}

\begin{equation}
\label{eq:scaling}
P(dt) = R  f(R dt) , \nonumber
\end{equation}

Therefore we evaluate the average rate for each distribution, as the inverse of the mean 
inter-arrival time $R=1/<dt>$, and rescale the inter-arrival time by the average rate. We 
find that the different distributions collapse quite well onto a universal curve 
(Fig. ref{poissDiffKscaled}), if the total number of links in the network is large enough. 
Small deviations are observed only for large $dt$. If the network is too sparse, i.e. if 
$\langle k \rangle \leq 3$, the mean inter-arrival time increases and the probability of 
longer $dt$ becomes larger. This effect is caused by the fact that in sparse networks only a 
small number of messages reaches the target. The space of possible paths leading to it is 
not fully explored and the trapping regions in sparse networks are more prominent.

\begin{figure}[t!]
\includegraphics*[width=0.45\textwidth]{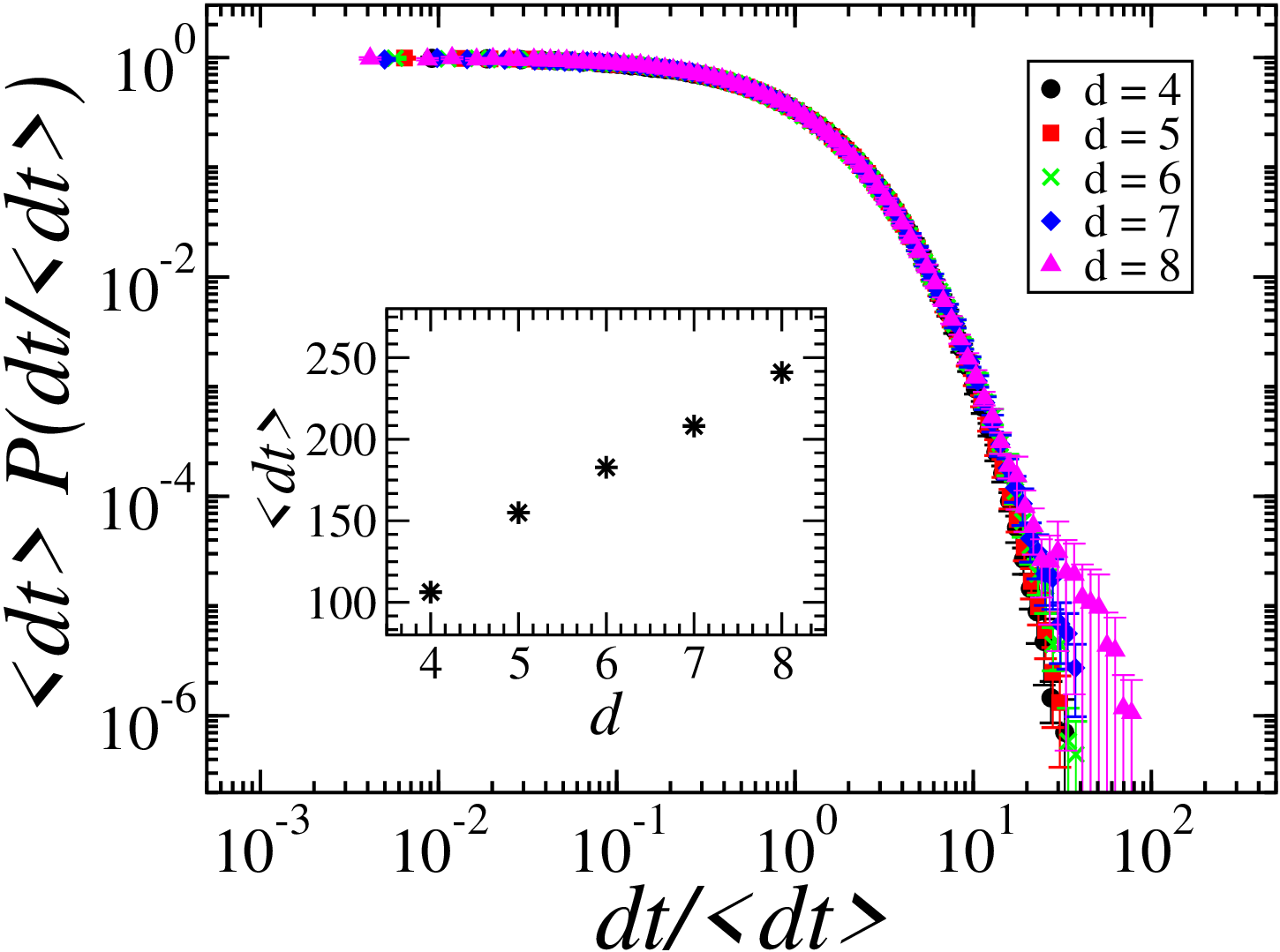}
\caption{
Distributions of inter-arrival times in networks with Poisson degree distribution with mean degree equal to 4 and different distances between the sender and the receiver. The distributions are rescaled by the average rate of message arrival. The networks have $N=10^4$ nodes, the number of sent messages is $M=10^6$. The messages are sent one by one. The inset shows the dependence of the mean inter-arrival time on the distance between the sender and the receiver.
}
\label{poissDiffDscaled}
\end{figure}

The universal scaling function in Fig.\ref{poissDiffKscaled} 
exhibits an initial almost constant regime followed by an exponential like decay. If the messages would arrive to the target independently of each other, inter-arrival times would be distributed exponentially, as it happens in Poissonian processes. In our process the distribution of inter-arrival times deviates from the exponential, which indicates that the process is more complex, and possibly correlations are present, coming from the networks topology or the message sending process itself.
We have also studied the influence of the distance $d$ between the sender and the receiver on the distribution of inter-arrival times. If the nodes are not too far away (for $d<8$), the rescaling of the distributions by the average rate of message arrivals provides a good collapse (Fig.\ref{poissDiffDscaled}) with fluctuations at large intertimes only for $d=8$. The inset shows that the mean inter-arrival time grows linearly with the sender - receiver distance.

\begin{figure}
\includegraphics*[width=0.45\textwidth]{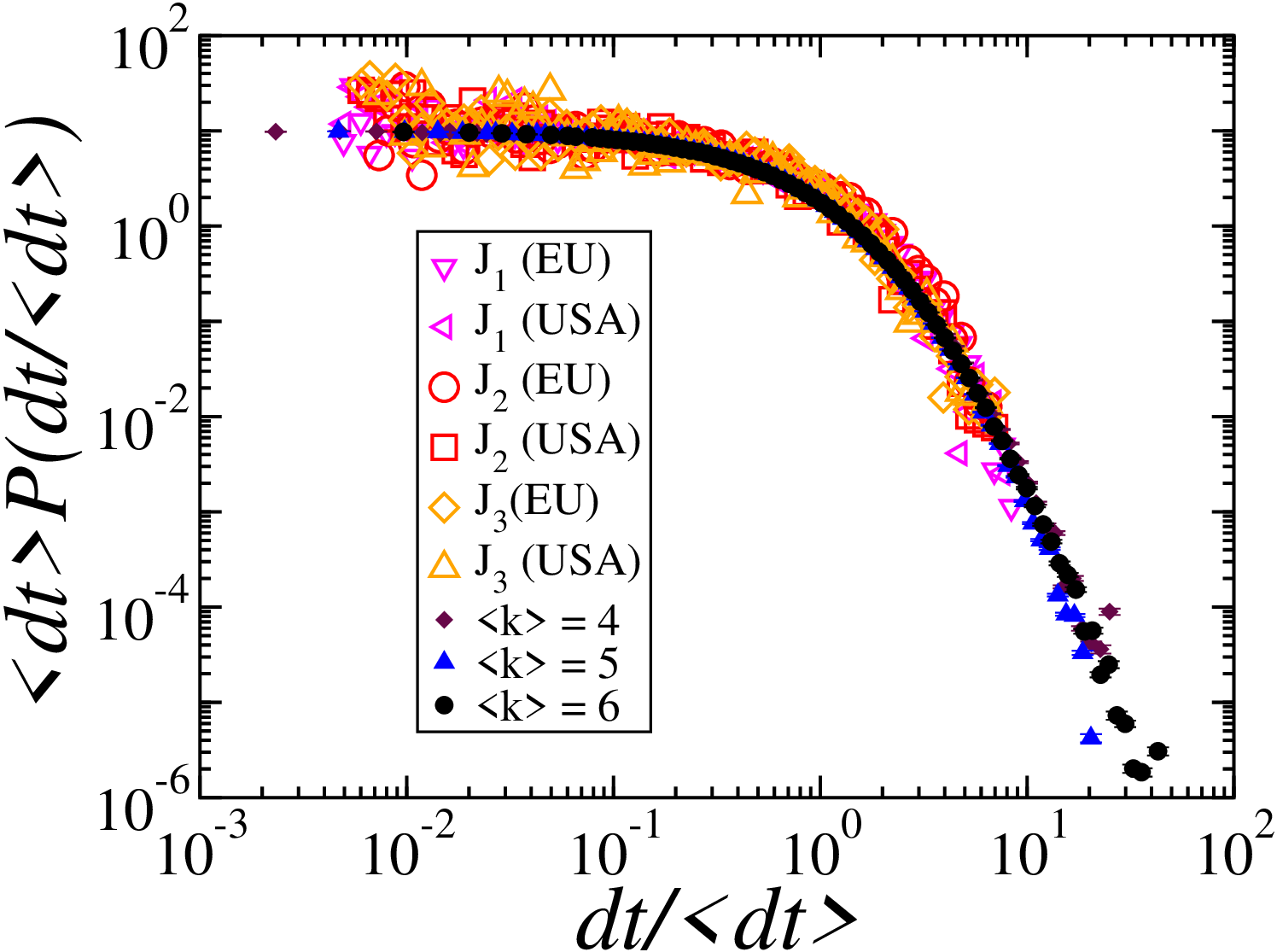}
\caption{Rescaled distributions of inter-arrival times of SPAM emails sent from two different domains and sampled in three different junk mailboxes (empty symbols) from reference \cite{lucillaSpam}, and of our model (full symbols) with messages sequentially sent  through random networks with different $\langle k \rangle$  values and the distance between the sender and the receiver fixed to 8. The networks have $N=10^4$ nodes, the number of sent messages is $M=10^6$.}
\label{spamData1}
\end{figure}

The present model simulates propagation of messages in unstructured peer to peer botnets where the botmaster is sequentially sending a large number of messages to the bots. The dynamics of 
sending orders inside the botnets influences the dynamics of arrival of messages sent from 
bots to the final destination, which could be a computer of a simple user receiving SPAM. 
Therefore, the fingerprint of the dynamics of message propagation inside the botnets should 
be visible in the distribution of inter-arrival times of SPAM emails collected in the mailbox 
of a single user. We compare our results with the distribution of inter-arrival times for 
SPAM data, presented in Ref. \cite{lucillaSpam}. The experimental data in 
 Fig.\ref{spamData1} are sampled from three different junk mailboxes, and the spam emails 
are selected on the basis of their geographical origin, Europe and the United States. 
Surprisingly, already our simple model is able to reproduce quite well the basic 
characteristics of the inter-arrival time distribution of the real data, as can be seen in 
Fig.\ref{spamData1}. 

\begin{figure}
\includegraphics*[width=0.45\textwidth]{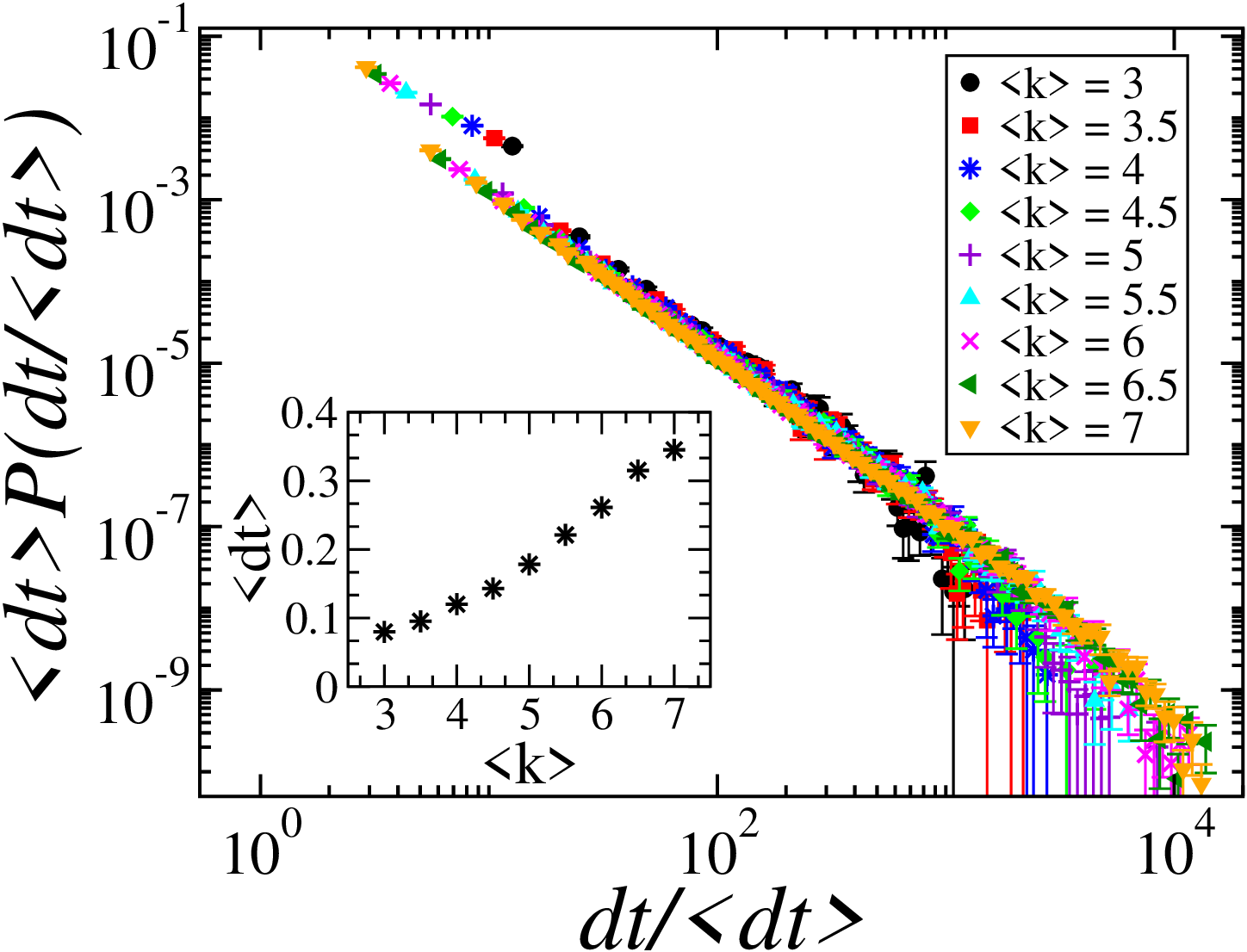}
\caption{
The rescaled distributions of inter-arrival times in networks with Poisson degree distribution with different mean degrees. All the messages are sent at the same time and the distance between the sender and the receiver is equal to 8. The networks have $N=10^4$ nodes, the number of sent messages is $M=10^6$.
}
\label{poissDpScaled}
\end{figure}

In order to understand the characteristics of the dynamical process, and thus the behavior 
of the inter-arrival time distribution for experimental data, we study in detail different 
aspects of the dynamics on model networks. If the walkers are sent one after the other the inter-arrival time does not depend only on the different paths taken by the walkers but also on their starting time. To understand the effect of this time delay on the process we also analyze the case where all the messages are sent at the same time. In this case inter-arrival times are uniquely determined by the complexity of the undertaken paths.
The distribution of inter-arrival times rescaled by the average rate shows universal behavior,
 well fitted by a power law with the exponent close to 2 (Fig.\ref{poissDpScaled}). We show 
results for only the networks which are not too sparse, since for sparse networks the 
distribution shows large statistical fluctuations. The power law behavior suggests that even 
if the walkers are completely independent during their propagation, they arrive in bursts 
originating temporal clustering in the process. The probability for the shortest $dt$ depends 
on the level of connectivity in the network, moreover the higher the average degree the 
smaller $dt$ is with respect to the average rate. 
In general, we observe that the average degree controls the extension of the scaling regime: the more inter-connected is the network the wider is the variety of possible paths and therefore the range of observed $dt$.

\begin{figure}
\includegraphics*[width=0.45\textwidth]{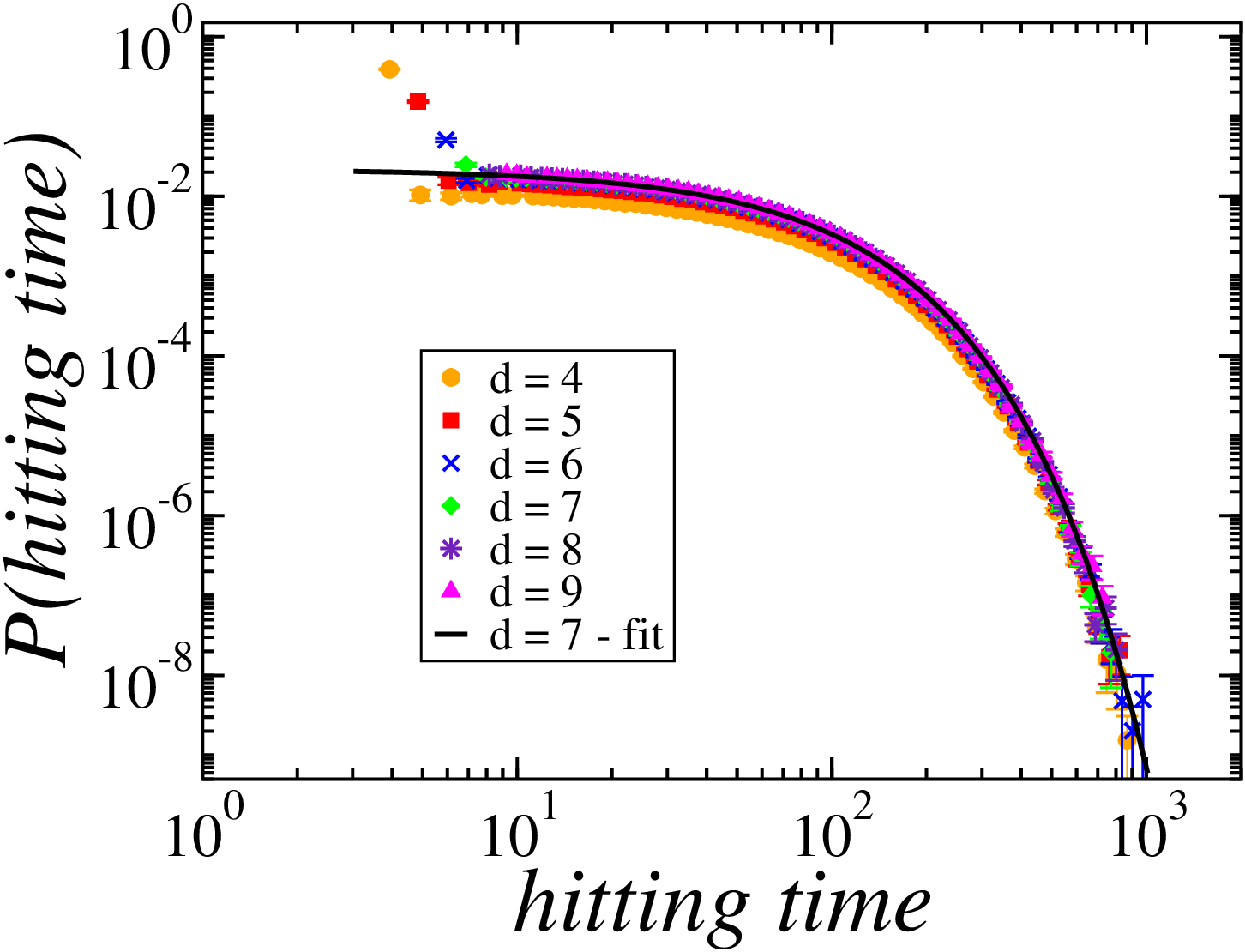}
\caption{
The distributions of hitting times in networks with Poisson degree distribution with mean degree equal to 4 and different distances between the sender and the receiver. The networks have $N=10^4$ nodes, the number of sent messages is $M=10^6$. The messages are sent all at once. The fit is a stretched exponential function of the form $y=0.02\cdot exp(-0.02\cdot x^{0.96})$ . }
\label{hittingTime}
\end{figure}

Since the outputs of random networks are distributed according to the Poisson degree 
distribution, a fraction of nodes in the network has $k=0$ outputs. Such nodes exist also in 
real networks. These nodes serve as a trap for the random walker. Similarly to the trapping 
problem on networks \cite{shlomoTrapping}, the message reaching such a node cannot proceed 
any farther. In random networks this is a dominant mechanism for preventing messages from 
finding the target. A large number of messages gets lost and the inter-arrival times can 
become extremely large. It is highly unlikely that a message will be stopped because the length of its path has reached the limit given by TTL. It rather appears that on random networks messages either reach the target after a relatively short period of time, or they never reach  it. We check this point by studying the distribution of hitting times. The hitting time, or the first passage time, is the time that a random walker takes to reach the target for the first time. In our model the distribution of hitting times is equivalent to the distribution of the lengths of the paths taken by the random walkers to reach the target node. In Fig.\ref{hittingTime} we see that only a small number of walkers takes the maximal number of 1000 steps, which is relatively small for the $N=10^4$ networks. 
The most probable hitting time has a value equal to the distance between the sender and the
receiver, meaning 
that many messages take the shortest path between the two nodes. Its probability is higher 
for smaller distances. The other possible paths are distributed according to a stretched 
exponential distribution (Fig.\ref{hittingTime}) independent of the sender - receiver 
distance.

\section{Random networks without dead ends}

In order to better understand the influence of nodes without outputs on the distribution of inter-arrival times, we study the same dynamical process on networks with slightly different topology. When we assign the number of ingoing or outgoing links to a node, we choose a random number between zero and $N-1$ from the Poisson distribution, but we assign to the node this number plus one. In this way there are no nodes with zero ingoing or outgoing links and there are no dead ends in the network. The only trap in the network is now the target node.

\begin{figure}
\includegraphics*[width=0.45\textwidth]{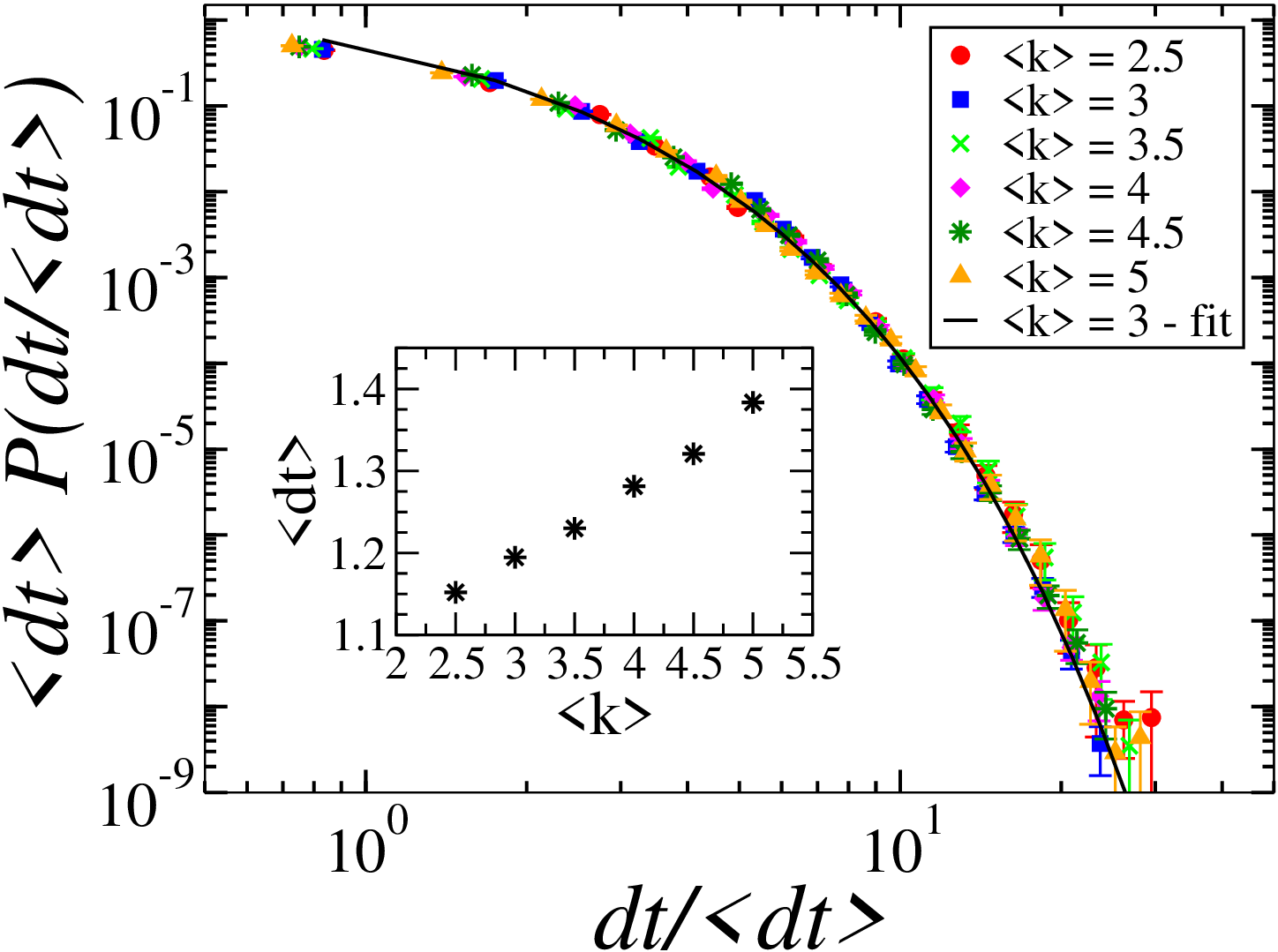}
\caption{
The distribution of inter-arrival times in networks with Poisson degree distribution without dead ends and with different values of the mean degree, rescaled by the average rate of message arrival. The networks have $N=10^4$ nodes, the number of sent messages is $M=10^6$ and the distance between the sender and the receiver is fixed equal to 5.  The messages are sent one by one. The fit is a stretched exponential function of the form $y=2.2\cdot exp(-1.5\cdot x^{0.8})$ . The insert shows the dependence of the mean inter-arrival time on the average degree of the network. 
}
\label{p1diffKscaled}
\end{figure}

\begin{figure}
\includegraphics*[width=0.45\textwidth]{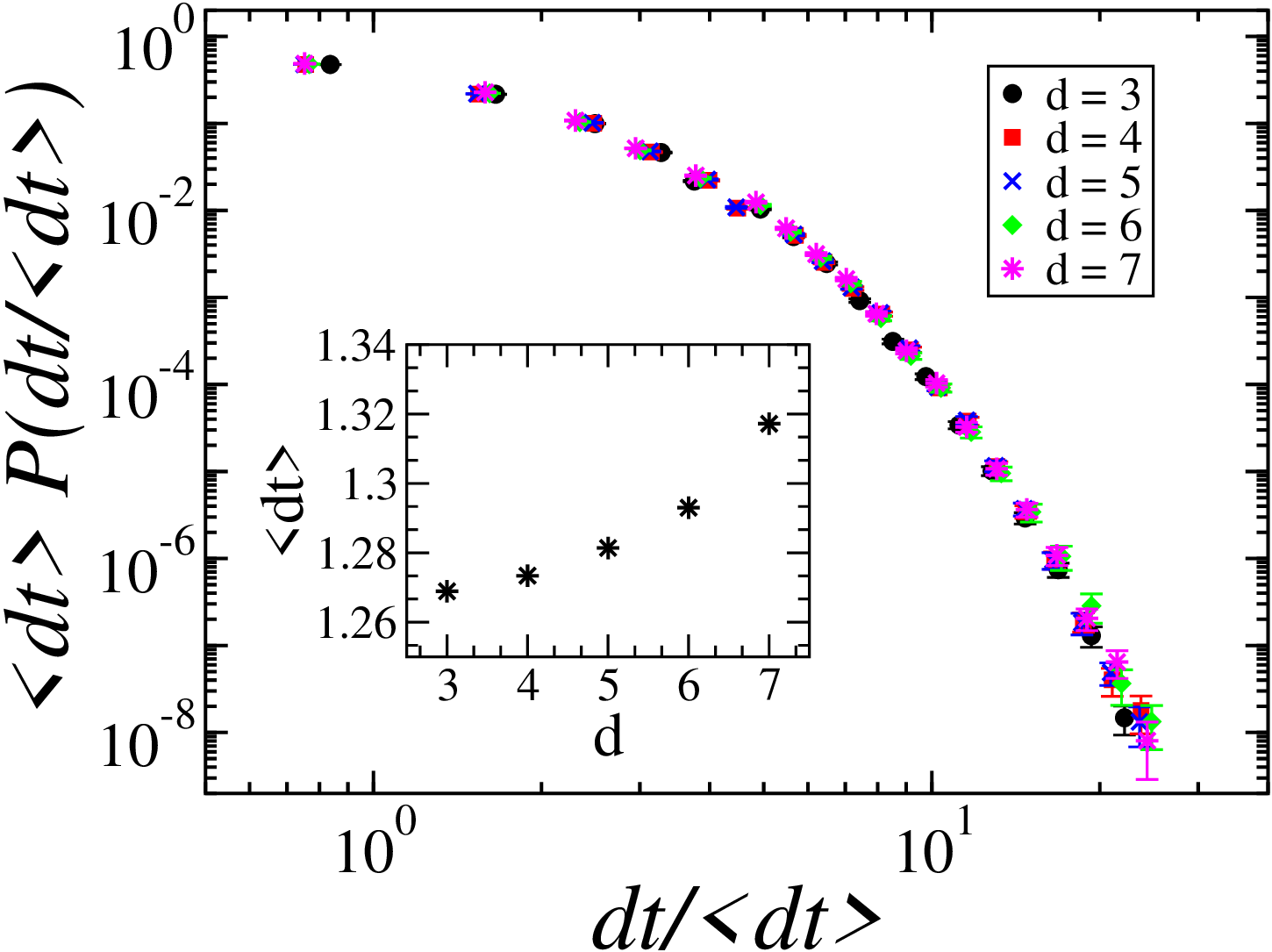}
\caption{
The distribution of inter-arrival times in networks with Poisson degree distribution without dead ends and with different distances between the sender and the receiver, rescaled by the average rate of message arrivals. The mean of the Poissonian distribution is $\langle k \rangle=4$. The networks have $N=10^4$ nodes, the number of sent messages is $M=10^6$ and the distance between the sender and the receiver is fixed equal to 5.  The messages are sent one by one. The inset shows the dependence of the mean inter-arrival time from the distance between the sender and the receiver. 
}
\label{p1diffDscaled}
\end{figure}

For the process where walkers are sent one by one we find that the distributions depend weekly on the average degree and the distance between the sender and the receiver. Indeed, the average rate varies on a much smaller range (insets in Fig.s\ref{p1diffKscaled} and \ref{p1diffDscaled}). Very good collapse is therefore observed rescaling the distributions by the average rate (Fig.s\ref{p1diffKscaled} and \ref{p1diffDscaled}). The universal function behaves as a stretched exponential and is therefore different than the one in Fig.\ref{poissDiffKscaled}. The main mechanism preventing the message from arriving to the target is now time exceeding the TTL limit. Many messages arrive to the target, which results in small inter-arrival times. In this case the walkers explore most of the paths existing between the two nodes. Similarly to previous cases, we show the results only for networks which are not too sparse, where the distributions show larger statistical fluctuations.

Conversely, for the process where all messages are sent at once, the inter-arrival times are 
in the majority of cases either zero, i.e. two messages arrive to the target at the same time,
 or equal to one. This is due to the large number of messages arriving to the target and to 
the ability of the walkers to explore well the space of all possible paths, with lengths 
ranging between the shortest path and the TTL.  At each time step then at least one walker 
arrives to the target leading to an inter-arrival time equal to one.

\begin{figure}
\includegraphics*[width=0.45\textwidth]{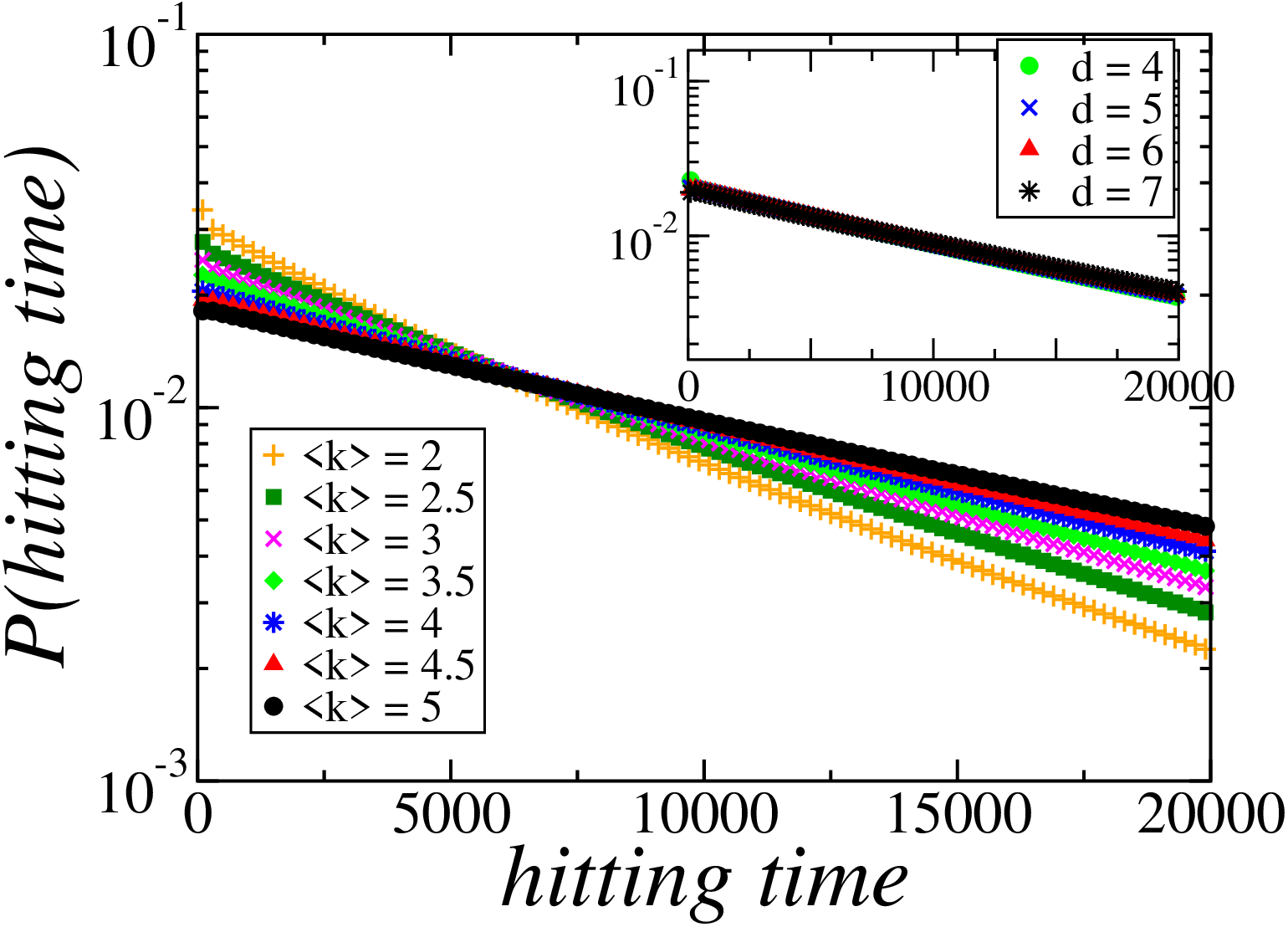}
\caption{
The distributions of hitting times in networks with Poisson degree distribution without dead ends and for different $\langle k \rangle$. The distance between the sender and the receiver is equal to 5. The inset shows the same distribution for the average degree equal to 4 and different distances between the sender and the receiver. The networks have $N=10^4$ nodes, the number of sent messages is $M=10^6$. }
\label{p1hittingTime}
\end{figure}

The distribution of hitting times when all messages are sent at once is also quite different 
than in the case where dead ends exist. In this case the walkers have the possibility to 
sample paths of all lengths and therefore the hitting time can assume values up to the 
threshold TTL. As we can see in Fig.\ref{p1hittingTime}, the distribution has an exponential 
behavior. Deviations from the exponential function can be seen only for small values of the 
hitting time, with this region getting smaller when the connectivity is larger. The 
coefficient of the exponential distribution increases with $\langle k \rangle$, namely the 
probability for longer hitting times is higher for larger average degrees. The walker takes 
more tortuous paths in a network with a higher level of connectivity. Conversely, the 
coefficient is independent of the distance between the sender and the receiver (inset 
Fig.\ref{p1hittingTime}).

The results obtained for random networks without dead ends confirm our conclusion that the 
behavior of the inter-arrival time distribution for random networks is a consequence of the 
existence of dead ends, the nodes without outputs which serve as traps for the messages. 
Since these traps exist in real networks, it is important to understand their influence on 
propagation of messages through random networks.

\section{Scale-free networks}

To explore further the influence of the network topology on the propagation of random walkers, we study this process on scale free networks. We find that in this case the characteristics of the process are much different. In order to measure the distribution we wait for the process to become stationary. Whereas in the case of Poisson distributed links this happens very fast, for scale free networks the stationary regime is reached after a long transient. A large number of messages has to be sent to obtain good statistics. 

 \begin{figure}
\includegraphics*[width=0.45\textwidth]{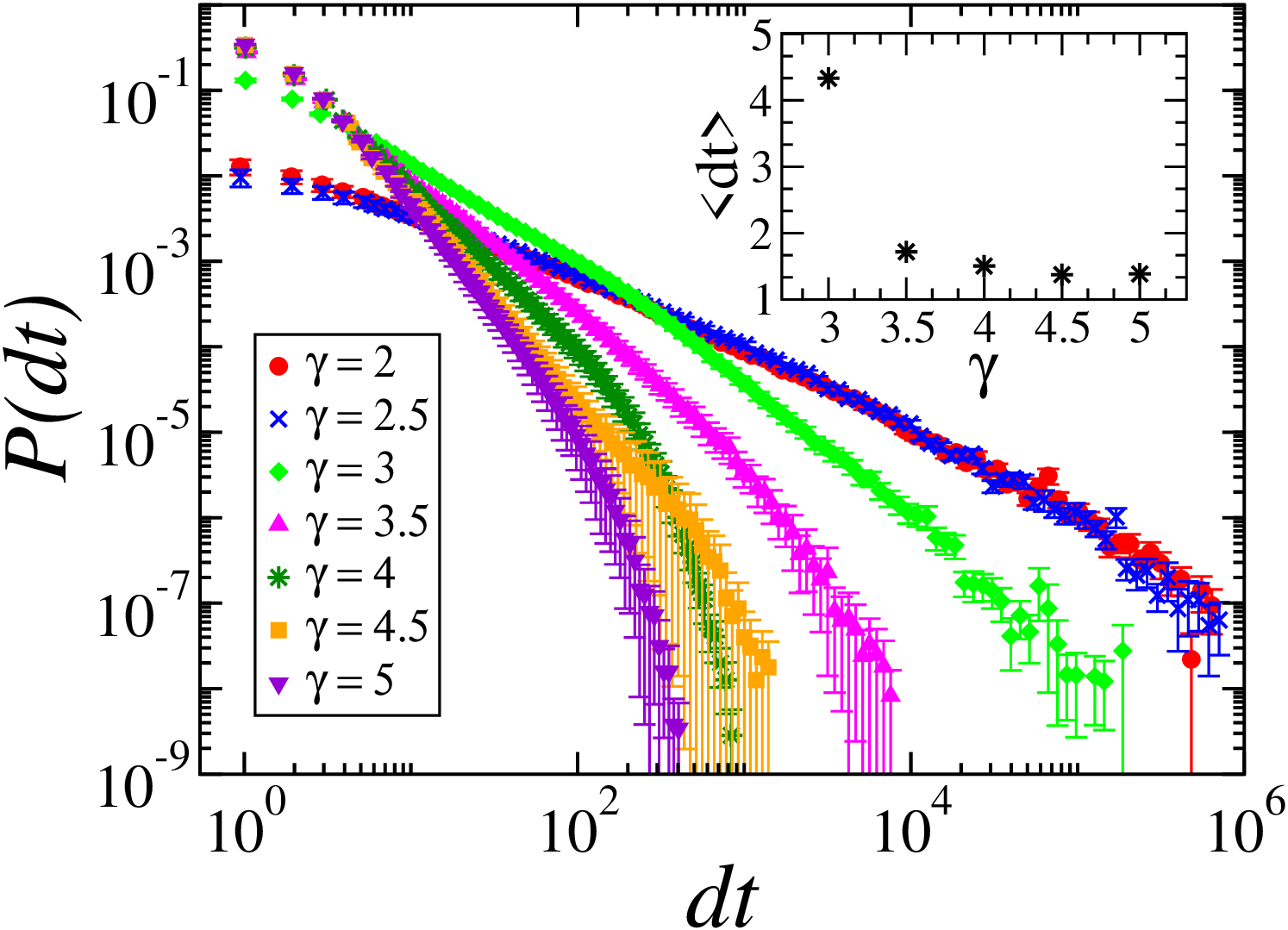}
\caption{
The distribution of inter-arrival times in networks with power law degree distribution  with different values of coefficient $\gamma$ ($P(k)\propto k^{-\gamma}$). The networks have $N=10^4$ nodes, the number of sent messages is $M=10^6$ and the distance between the sender and the receiver is fixed equal to 30. The messages are sent one by one. The inset shows the dependence of the mean inter-arrival time on $\gamma$. 
}
\label{sfDiffGamma}
\end{figure}

In Fig.\ref{sfDiffGamma} we show the distribution of inter-arrival times for scale-free networks with different exponents of the degree distribution ($P(k)\propto k^{-\gamma}$) and in the case that messages are sent one by one. We observe that sparse networks (high $\gamma$) behave differently than well connected networks. In fact, the mean inter-arrival time rapidly increases for decreasing $\gamma$, suggesting that the walker takes very tortuous paths in a well connected network. In contrast to the previous cases, the distributions of inter-arrival times for scale free networks do not collapse onto a universal curve if the inter-arrival time is rescaled by the average rate of the process, even when the networks are not sparse.
Therefore the average rate is not the only relevant quantity in the process.

The number of messages arriving to the target is smaller than in Poisson networks without dead ends, and the inter-arrival times can be extremely long. In scale free networks there are by definition no dead ends, and the main mechanism for stopping the random walker is here time exceeding the TTL limit. The walkers take many long paths, probably looping through system and being able to visit different parts of the networks through shortcuts whose probability is higher due to hubs. In contrast to Poisson networks without dead ends, although the walkers can explore well the space of possible paths, many of them never reach the target since they are either trapped in loops, or in regions typical for directed networks, where the walker can enter but cannot escape from. Moreover, the path to the target could be longer than the TTL limit, which is very probable on scale free networks. 
Since the number of walkers reaching the target decreases, the probability of longer 
inter-arrival times increases.

\begin{figure}
\includegraphics*[width=0.45\textwidth]{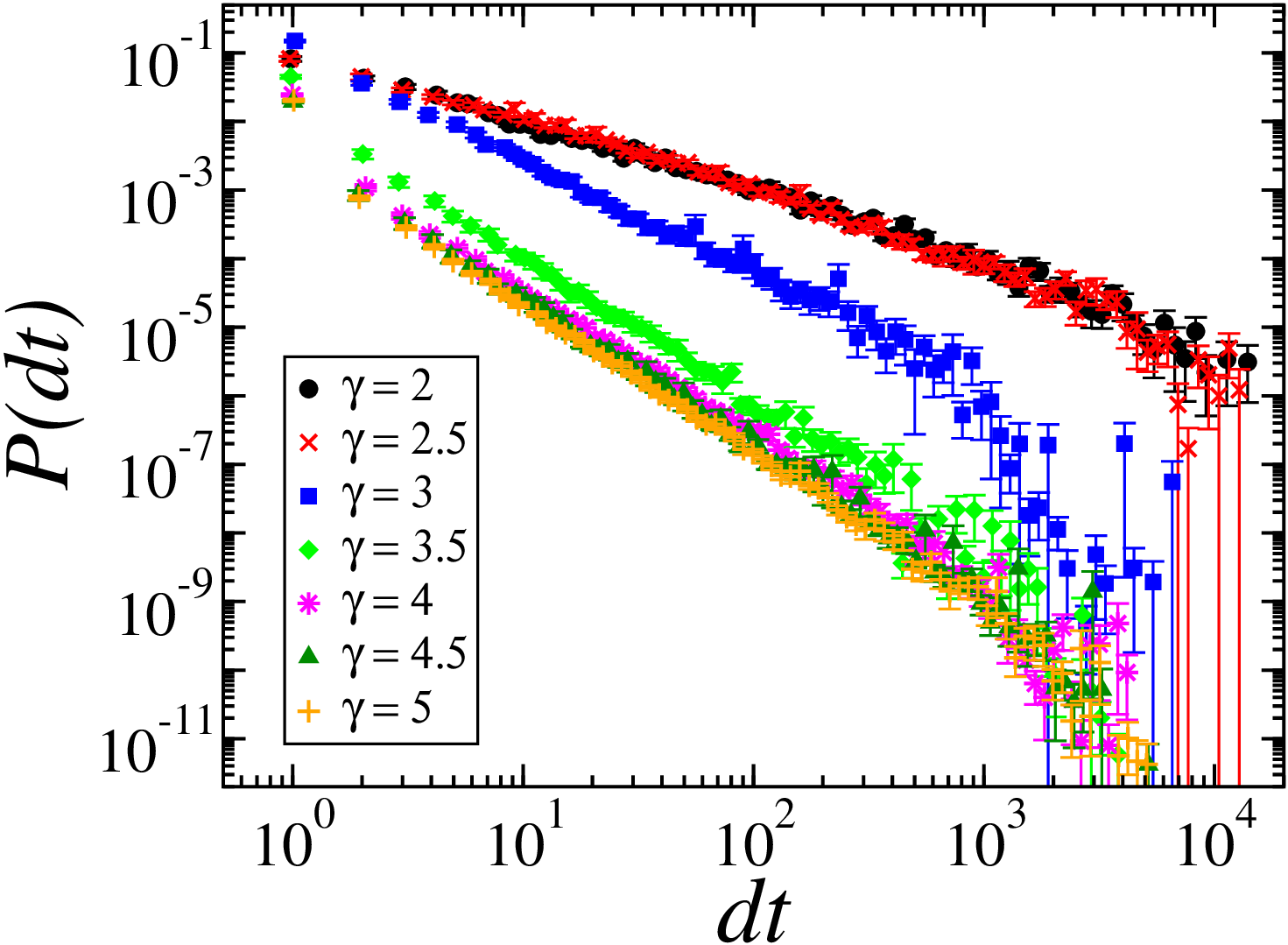}
\caption{
The distributions of inter-arrival times in networks with scale free distribution with different coefficients of the power law. All the messages are sent at the same time and the distance between the sender and the receiver is 30. The networks have $N=10^4$ nodes, the number of sent messages is $M=10^6$. 
}
\label{sfDiffGammaDp}
\end{figure}

When messages are sent all at the same time, the distribution of inter-arrival times shows a 
power law behavior. However, in contrast to random networks with dead ends, where we find 
universal power law behavior, a change in the topology by tuning the coefficient $\gamma$ 
changes the slope from about 0.8, in well connected, to 2.2, in sparse networks 
(Fig.\ref{sfDiffGammaDp}). As in the case of consecutive departures of messages, also here we 
do not observe the distribution collapse if  inter-arrival time is rescaled by the average 
rate.

\begin{figure}[t!]
\includegraphics*[width=0.45\textwidth]{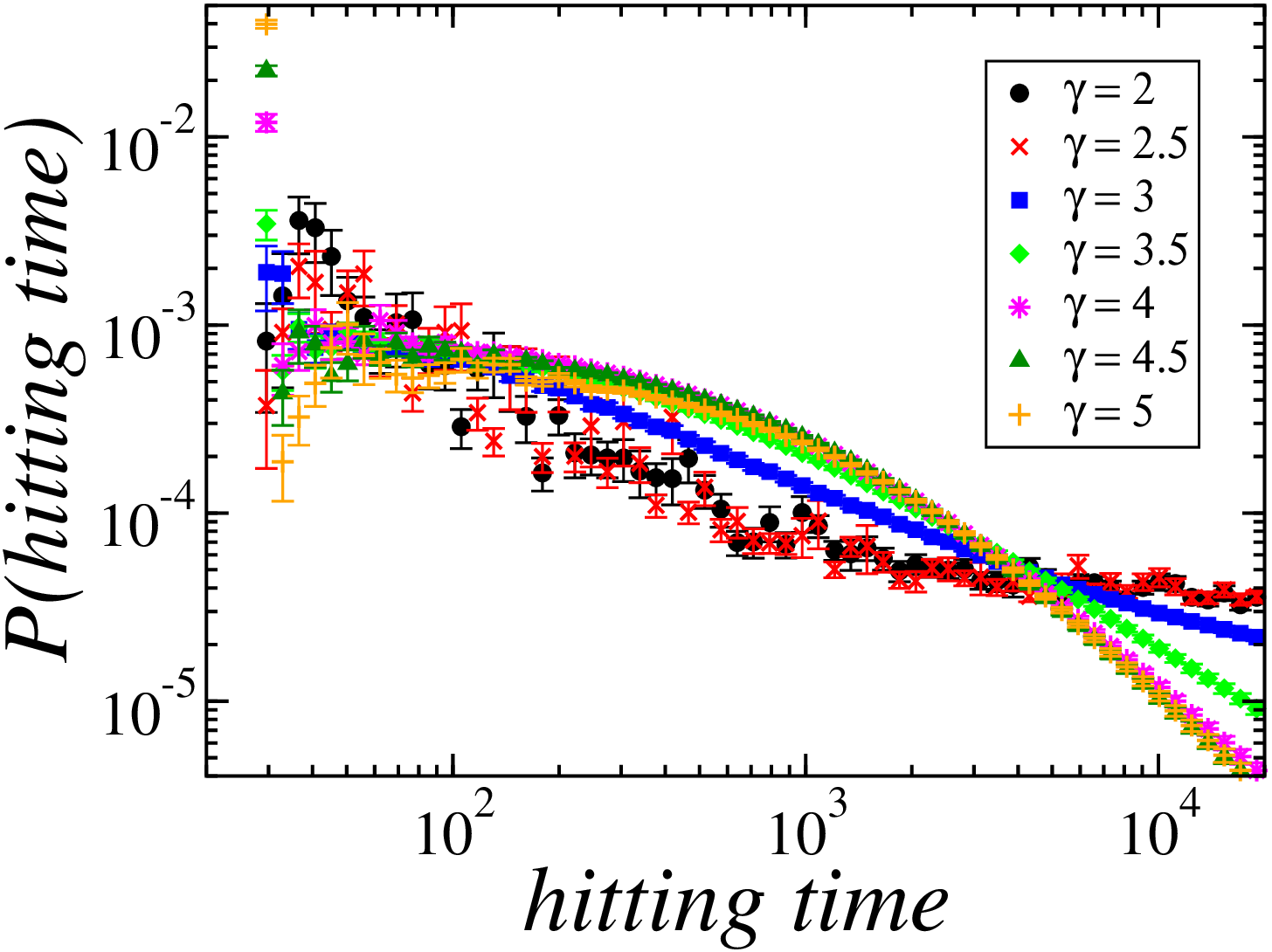}
\caption{
The distributions of hitting times in networks with power law degree distributions with different coefficients $\gamma$. The distance between the sender and the receiver is fixed to 30. The networks have $N=10^4$ nodes, the number of sent messages is $M=10^6$.}
\label{sfHittingTime}
\end{figure}

The hitting times of the dynamical process on scale free networks also show a quite different
behavior. In Fig.\ref{sfHittingTime} we see that networks with different power law 
coefficients have different distributions of hitting times. We also see that, similarly to the case of Poisson networks without dead ends, the walkers on scale free networks explore well the space of all possible paths from shortest path to TTL, especially in the case of well connected networks (smaller $\gamma$ exponents). By increasing $\gamma$, the number of walkers which take very long paths to reach the target decreases.

\section{Conclusions}

In this paper we study the distribution of inter-arrival times of walkers sent through 
complex networks, with the goal to gain understanding on the process of information spreading 
inside the new generation of botnets based on peer to peer communication. The dynamics of 
sending botmaster orders inside the botnet influences the dynamics of arrival of messages 
sent from bots to the final destination. This can be a computer of a simple user receiving 
either SPAM or data packages contaminated with viruses. The dynamics of message propagation 
inside the botnets  therefore affects the distribution of inter-arrival 
times of SPAM emails collected in the mailbox of a single user. We compare the results 
obtained by modeling botnets as random directed networks, where messages sent sequentially 
from the botmaster are  random walkers, with the distribution of inter-arrival times of real 
SPAM data, presented in the Ref. \cite{lucillaSpam}. The comparison shows that this simple 
model reproduces well the basic features of inter-arrival time distribution.

To better understand the behavior of the distribution of the inter-arrival times  we study the dynamical process of message propagation on different model networks and by two different procedures. We find that the main ingredients controlling the distribution of inter-arrival times are the distribution of possible path lengths between the sender and the receiver, and the number of messages not reaching the target. Possible paths between two nodes are determined only by the network topology while the mechanisms preventing messages from reaching the target depend in addition on TTL. In the case of random networks, nodes without outputs represent natural 
traps for the messages, and most of the sent messages are prevented from reaching the target 
leading to long inter-arrival times. Since such nodes exist in real networks, this is the situation that we expect to observe in real peer to peer botnets. The distribution of inter-arrival times for the sequential sending of messages shows an almost constant regime for small 
inter-arrival times, followed by an exponential like cutoff including non-vanishing 
probability for very long inter-arrival times. When the messages are sent in parallel the 
inter-arrival times are power law distributed up to long inter-arrival times.

We confirm that the nodes without outputs have a crucial role in processes on random networks 
by studying networks with the same Poisson distribution of links, but without dead ends. In 
this case inter-arrival times are much shorter and are distributed as a stretched exponential 
for sequentially sent messages. When the messages are sent in parallel  only trivial 
values of inter-arrival times, zero or one, appear. Many messages arrive to the target, the 
space of possible paths is well explored and only those few messages exceeding the limit of 
maximal number of steps are prevented from reaching the target.

Networks with a scale free distribution of links also show the important role of the network 
topology for the behavior of the inter-arrival time distribution. In these networks long inter-arrival times appear, but the messages are prevented from reaching the target by different mechanisms. Here the limit on the maximal number of steps, together with the distribution of possible paths between the sender and the receiver, are determining the distributions of inter-arrival times. The inter-arrival times of sequentially sent messages can be very long in the case of well connected networks, or much shorter for the less connected networks (higher $\gamma$ values), but the behaviour of the distributions are in all cases different than for the other two 
network types. For messages sent in parallel this distribution is a power law, as in the case 
of random networks, but with the slope depending on the network's connectivity. From the 
three types of networks investigated, only for scale free networks the distributions do not 
collapse onto a universal curve if $dt$ is rescaled by the average rate of the process. 

The change in topology has a different influence on the dynamical process on networks with Poisson and scale free distributed links. For scale free networks we change the topology by varying the $\gamma$ exponent of the link distribution. This change has a strong influence on the 
distribution of message inter-arrival times. The average rate is not the only relevant 
quantity for the process and the universality class of the distribution of inter-arrival times 
depends on the topology. Tuning the exponent $\gamma$ affects the number of hubs in the 
network, which in scale free networks play a crucial role in the process of message 
spreading. Their number changes qualitatively the behavior of the inter-arrival time 
distribution. It influences the length of possible paths between nodes in the network not 
only by the change of local properties, such as the number of links of a node, but also 
by the creation of long range shortcuts through hubs, which increases the number of possible 
paths between the nodes.

When the link distribution is Poissonian, changing the mean 
number of links per node modifies the network topology and therefore the mean rate of walker 
arrivals. 
The inter-arrival time distributions, however, collapse onto a universal scaling function 
if inter-arrival time is rescaled by the average rate. The distance between the sender and 
the receiver is also not affecting the universality class of the universal feature of the 
distribution. The behavior of the distribution is also robust with respect to changes of 
other parameters, such as the time distance between two sequential messages, the number of 
output links of the sender and input links of the receiver, or even the number of senders. 
The robustness of the behavior of the inter-arrival time distribution seems to be typical 
for networks with Poisson distributed links. Interestingly, the SPAM data, analyzed in terms 
of the junk mailbox, or by the geographical location of IP addresses of the sender, also 
collapse onto a universal function when inter-arrival time is rescaled by the average rate.
This new approach in studying such processes using network theory can be 
employed in many fields. By applying this approach to directed networks we show that we can 
learn about 
botnets indirectly. The results are not botnet specific and can be applied to 
any other system which can be modeled by directed networks and through which the information 
propagates in random fashion. 

We acknowledge financial support from the ETH Competence Center 'Coping with Crises in Complex Socio-Economic Systems' (CCSS) through ETH Research Grant CH1-01-08-2, the financial support from the Swiss National Science Foundation (grant number 200021-126853) and FUNCAP.


\begin{thebibliography}{40}
\expandafter\ifx\csname natexlab\endcsname\relax\def\natexlab#1{#1}\fi
\expandafter\ifx\csname bibnamefont\endcsname\relax
  \def\bibnamefont#1{#1}\fi
\expandafter\ifx\csname bibfnamefont\endcsname\relax
  \def\bibfnamefont#1{#1}\fi
\expandafter\ifx\csname citenamefont\endcsname\relax
  \def\citenamefont#1{#1}\fi
\expandafter\ifx\csname url\endcsname\relax
  \def\url#1{\texttt{#1}}\fi
\expandafter\ifx\csname urlprefix\endcsname\relax\def\urlprefix{URL }\fi
\providecommand{\bibinfo}[2]{#2}
\providecommand{\eprint}[2][]{\url{#2}}

\bibitem{eu} European Union press release, reference IP/03/101
15/07/2003.

\bibitem[{\citenamefont{Grizzard et~al.}(2007)\citenamefont{Grizzard, Sharma,
  Nunnery, Kang, and Dagon}}]{p2pBotnets}
\bibinfo{author}{\bibfnamefont{J.~B.} \bibnamefont{Grizzard}},
  \bibinfo{author}{\bibfnamefont{V.}~\bibnamefont{Sharma}},
  \bibinfo{author}{\bibfnamefont{C.}~\bibnamefont{Nunnery}},
  \bibinfo{author}{\bibfnamefont{B.~B.} \bibnamefont{Kang}}, \bibnamefont{and}
  \bibinfo{author}{\bibfnamefont{D.}~\bibnamefont{Dagon}},
  \bibinfo{journal}{Proceedings of the first conference on Hot Topics in
  Understanding Botnets} p.~\bibinfo{pages}{1} (\bibinfo{year}{2007}).

\bibitem[{\citenamefont{Bisnik and Abouzeid}(2005)}]{otherP2P}
\bibinfo{author}{\bibfnamefont{N.}~\bibnamefont{Bisnik}} \bibnamefont{and}
  \bibinfo{author}{\bibfnamefont{A.~A.} \bibnamefont{Abouzeid}},
  \bibinfo{journal}{Proceedings of the Second International Workshop on Hot
  Topics in Peer-to-Peer Systems} p.~\bibinfo{pages}{95}
  (\bibinfo{year}{2005}).

\bibitem[{\citenamefont{Li and Zhang}(2010)}]{directedWirelessRouting}
\bibinfo{author}{\bibfnamefont{Y.}~\bibnamefont{Li}} \bibnamefont{and}
  \bibinfo{author}{\bibfnamefont{Z.}~\bibnamefont{Zhang}},
  \bibinfo{journal}{INFCOM, 2010 Proceedings IEEE} p.~\bibinfo{pages}{1}
  (\bibinfo{year}{2010}).

\bibitem[{\citenamefont{Bar-Yossef et~al.}(2006)\citenamefont{Bar-Yossef, R,
  and Kliot}}]{adHocNets}
\bibinfo{author}{\bibfnamefont{Z.}~\bibnamefont{Bar-Yossef}},
  \bibinfo{author}{\bibfnamefont{F.}~\bibnamefont{R}}, \bibnamefont{and}
  \bibinfo{author}{\bibfnamefont{G.}~\bibnamefont{Kliot}},
  \bibinfo{journal}{Proceedings of the seventh ACM International Symposium on
  Mobile ad hoc Networking and Computing} p. \bibinfo{pages}{238}
  (\bibinfo{year}{2006}).

\bibitem[{\citenamefont{Capocci et~al.}(2009)\citenamefont{Capocci,
  Baldassarri, Servedio, and Loreto}}]{loretoTagging}
\bibinfo{author}{\bibfnamefont{A.}~\bibnamefont{Capocci}},
  \bibinfo{author}{\bibfnamefont{A.}~\bibnamefont{Baldassarri}},
  \bibinfo{author}{\bibfnamefont{V.~D.~P.} \bibnamefont{Servedio}},
  \bibnamefont{and} \bibinfo{author}{\bibfnamefont{V.}~\bibnamefont{Loreto}},
  \bibinfo{journal}{Proceedings of the 20th ACM conference on Hypertext and
  Hypermedia} p. \bibinfo{pages}{239} (\bibinfo{year}{2009}).

\bibitem[{\citenamefont{Hughes}(1995)}]{rwBook}
\bibinfo{author}{\bibfnamefont{B.~D.} \bibnamefont{Hughes}},
  \emph{\bibinfo{title}{Random Walks and Random Enviroments, Vol.1: Random
  Walks}} (\bibinfo{publisher}{Oxford University Press, New York},
  \bibinfo{year}{1995}).

\bibitem[{\citenamefont{Boccaletti et~al.}(2006)\citenamefont{Boccaletti,
  Latora, Moreno, Chavez, and Hwang}}]{boccaletti}
\bibinfo{author}{\bibfnamefont{S.}~\bibnamefont{Boccaletti}},
  \bibinfo{author}{\bibfnamefont{V.}~\bibnamefont{Latora}},
  \bibinfo{author}{\bibfnamefont{Y.}~\bibnamefont{Moreno}},
  \bibinfo{author}{\bibfnamefont{M.}~\bibnamefont{Chavez}}, \bibnamefont{and}
  \bibinfo{author}{\bibfnamefont{D.-U.} \bibnamefont{Hwang}},
  \bibinfo{journal}{Physics Reports} \textbf{\bibinfo{volume}{424}},
  \bibinfo{pages}{175} (\bibinfo{year}{2006}).

\bibitem[{\citenamefont{Barrat et~al.}(2008)\citenamefont{Barrat, Barthelemy,
  and Vespignani}}]{vespignani}
\bibinfo{author}{\bibfnamefont{A.}~\bibnamefont{Barrat}},
  \bibinfo{author}{\bibfnamefont{M.}~\bibnamefont{Barthelemy}},
  \bibnamefont{and}
  \bibinfo{author}{\bibfnamefont{A.}~\bibnamefont{Vespignani}},
  \emph{\bibinfo{title}{Dynamical processes in complex networks}}
  (\bibinfo{publisher}{Cambridge University Press}, \bibinfo{year}{2008}).

\bibitem[{\citenamefont{Noh and Rieger}(2004)}]{riegerRWonComplexNets}
\bibinfo{author}{\bibfnamefont{J.}~\bibnamefont{Noh}} \bibnamefont{and}
  \bibinfo{author}{\bibfnamefont{H.}~\bibnamefont{Rieger}},
  \bibinfo{journal}{Phys. Rev. Lett.} \textbf{\bibinfo{volume}{92}},
  \bibinfo{pages}{118701} (\bibinfo{year}{2004}).

\bibitem[{\citenamefont{Gallos}(2004)}]{gallosRandomTrapping}
\bibinfo{author}{\bibfnamefont{L.~K.} \bibnamefont{Gallos}},
  \bibinfo{journal}{Phys. Rev. E} \textbf{\bibinfo{volume}{70}},
  \bibinfo{pages}{046116} (\bibinfo{year}{2004}).

\bibitem[{\citenamefont{Condamin et~al.}(2007)\citenamefont{Condamin,
  B\'enichou, Tejedor, Voituriez, and Klafter}}]{klafterNature}
\bibinfo{author}{\bibfnamefont{S.}~\bibnamefont{Condamin}},
  \bibinfo{author}{\bibfnamefont{v.}~\bibnamefont{B\'enichou}},
  \bibinfo{author}{\bibfnamefont{V.}~\bibnamefont{Tejedor}},
  \bibinfo{author}{\bibfnamefont{J.}~\bibnamefont{Voituriez}},
  \bibnamefont{and} \bibinfo{author}{\bibfnamefont{J.}~\bibnamefont{Klafter}},
  \bibinfo{journal}{Nature} \textbf{\bibinfo{volume}{450}}, \bibinfo{pages}{77}
  (\bibinfo{year}{2007}).

\bibitem[{\citenamefont{Sood et~al.}(2005)\citenamefont{Sood, Redner, and ben
  Avraham}}]{rednerFPTerdosRenyi}
\bibinfo{author}{\bibfnamefont{V.}~\bibnamefont{Sood}},
  \bibinfo{author}{\bibfnamefont{S.}~\bibnamefont{Redner}}, \bibnamefont{and}
  \bibinfo{author}{\bibfnamefont{D.}~\bibnamefont{Ben Avraham}},
  \bibinfo{journal}{J. Phys. A : Math. Gen.} \textbf{\bibinfo{volume}{38}},
  \bibinfo{pages}{109} (\bibinfo{year}{2005}).

\bibitem[{\citenamefont{ben Avraham and Havlin}(2000)}]{shlomoBook}
\bibinfo{author}{\bibfnamefont{D.}~\bibnamefont{Ben Avraham}} \bibnamefont{and}
  \bibinfo{author}{\bibfnamefont{S.}~\bibnamefont{Havlin}},
  \emph{\bibinfo{title}{Diffusion and Reactions in Fractals and Disordered
  Systems}} (\bibinfo{publisher}{Cambrige University Press, New York},
  \bibinfo{year}{2000}).

\bibitem[{\citenamefont{Adamic et~al.}(2001)\citenamefont{Adamic, Lukose,
  Punliyani, and Huberman}}]{ladaAdamic}
\bibinfo{author}{\bibfnamefont{L.~A.} \bibnamefont{Adamic}},
  \bibinfo{author}{\bibfnamefont{A.}~\bibnamefont{Lukose}},
  \bibinfo{author}{\bibfnamefont{A.}~\bibnamefont{Punliyani}},
  \bibnamefont{and} \bibinfo{author}{\bibfnamefont{B.~A.}
  \bibnamefont{Huberman}}, \bibinfo{journal}{Phys. Rev. E}
  \textbf{\bibinfo{volume}{64}}, \bibinfo{pages}{046135}
  (\bibinfo{year}{2001}).

\bibitem[{\citenamefont{Kim et~al.}(2002)\citenamefont{Kim, Yoon, Han, and
  Jeong}}]{kim}
\bibinfo{author}{\bibfnamefont{B.}~\bibnamefont{Kim}},
  \bibinfo{author}{\bibfnamefont{C.}~\bibnamefont{Yoon}},
  \bibinfo{author}{\bibfnamefont{S.}~\bibnamefont{Han}}, \bibnamefont{and}
  \bibinfo{author}{\bibfnamefont{H.}~\bibnamefont{Jeong}},
  \bibinfo{journal}{Phys. Rev. E} \textbf{\bibinfo{volume}{65}},
  \bibinfo{pages}{027103} (\bibinfo{year}{2002}).

\bibitem[{\citenamefont{Rosvall et~al.}(2005)\citenamefont{Rosvall, Minnhagen,
  and Sneppen}}]{sneppen}
\bibinfo{author}{\bibfnamefont{M.}~\bibnamefont{Rosvall}},
  \bibinfo{author}{\bibfnamefont{P.}~\bibnamefont{Minnhagen}},
  \bibnamefont{and} \bibinfo{author}{\bibfnamefont{K.}~\bibnamefont{Sneppen}},
  \bibinfo{journal}{Phys. Rev. E} \textbf{\bibinfo{volume}{71}},
  \bibinfo{pages}{066111} (\bibinfo{year}{2005}).

\bibitem[{\citenamefont{Germano and de~Moura}(2006)}]{germano}
\bibinfo{author}{\bibfnamefont{R.}~\bibnamefont{Germano}} \bibnamefont{and}
  \bibinfo{author}{\bibfnamefont{A.~P.~S.} \bibnamefont{de~Moura}},
  \bibinfo{journal}{Phys. Rev. E} \textbf{\bibinfo{volume}{74}},
  \bibinfo{pages}{036117} (\bibinfo{year}{2006}).

\bibitem[{\citenamefont{Tadic and Rodgers}(2002)}]{bosaTransport}
\bibinfo{author}{\bibfnamefont{B.}~\bibnamefont{Tadic}} \bibnamefont{and}
  \bibinfo{author}{\bibfnamefont{J.}~\bibnamefont{Rodgers}},
  \bibinfo{journal}{Adv. Complex Syst.} \textbf{\bibinfo{volume}{5}},
  \bibinfo{pages}{445} (\bibinfo{year}{2002}).

\bibitem[{\citenamefont{Jaspersen and Blumen}(2000)}]{jaspersenBlumen}
\bibinfo{author}{\bibfnamefont{S.}~\bibnamefont{Jaspersen}} \bibnamefont{and}
  \bibinfo{author}{\bibfnamefont{A.}~\bibnamefont{Blumen}},
  \bibinfo{journal}{Phys. Rev. E} \textbf{\bibinfo{volume}{62}},
  \bibinfo{pages}{6270} (\bibinfo{year}{2000}).

\bibitem[{\citenamefont{Newman and Girvan}(2004)}]{newmanRWbetweeness}
\bibinfo{author}{\bibfnamefont{M.~E.~J.} \bibnamefont{Newman}}
  \bibnamefont{and} \bibinfo{author}{\bibfnamefont{M.}~\bibnamefont{Girvan}},
  \bibinfo{journal}{Phys. Rev. E} \textbf{\bibinfo{volume}{69}},
  \bibinfo{pages}{026113} (\bibinfo{year}{2004}).

\bibitem[{\citenamefont{Zhou}(2003{\natexlab{a}})}]{zhouRwCommDet1}
\bibinfo{author}{\bibfnamefont{H.}~\bibnamefont{Zhou}}, \bibinfo{journal}{Phys.
  Rev. E} \textbf{\bibinfo{volume}{67}}, \bibinfo{pages}{041908}
  (\bibinfo{year}{2003}{\natexlab{a}}).

\bibitem[{\citenamefont{Zhou}(2003{\natexlab{b}})}]{zhouRwCommDet2}
\bibinfo{author}{\bibfnamefont{H.}~\bibnamefont{Zhou}}, \bibinfo{journal}{Phys.
  Rev. E} \textbf{\bibinfo{volume}{67}}, \bibinfo{pages}{061901}
  (\bibinfo{year}{2003}{\natexlab{b}}).

\bibitem[{\citenamefont{Sood and Grassberger}(2007)}]{vishalRWbiased}
\bibinfo{author}{\bibfnamefont{V.}~\bibnamefont{Sood}} \bibnamefont{and}
  \bibinfo{author}{\bibfnamefont{P.}~\bibnamefont{Grassberger}},
  \bibinfo{journal}{Phys. Rev. Lett.} \textbf{\bibinfo{volume}{99}},
  \bibinfo{pages}{098701} (\bibinfo{year}{2007}).

\bibitem[{\citenamefont{Kim et~al.}(2010)\citenamefont{Kim, Son, and
  Jeong}}]{jeongDirectedCitationNets}
\bibinfo{author}{\bibfnamefont{Y.}~\bibnamefont{Kim}},
  \bibinfo{author}{\bibfnamefont{S.}~\bibnamefont{Son}}, \bibnamefont{and}
  \bibinfo{author}{\bibfnamefont{H.}~\bibnamefont{Jeong}},
  \bibinfo{journal}{Phys. Rev. E} \textbf{\bibinfo{volume}{81}},
  \bibinfo{pages}{016103} (\bibinfo{year}{2010}).

\bibitem[{\citenamefont{Tadic}(2001)}]{bosaDirected}
\bibinfo{author}{\bibfnamefont{B.}~\bibnamefont{Tadic}}, \bibinfo{journal}{Eur.
  Phys. J. B} \textbf{\bibinfo{volume}{23}}, \bibinfo{pages}{221}
  (\bibinfo{year}{2001}).

\bibitem[{\citenamefont{Fortunato and Flammini}(2007)}]{fortunatoPagerank}
\bibinfo{author}{\bibfnamefont{S.}~\bibnamefont{Fortunato}} \bibnamefont{and}
  \bibinfo{author}{\bibfnamefont{A.}~\bibnamefont{Flammini}},
  \bibinfo{journal}{Internat. J. Bifur. Chaos.} \textbf{\bibinfo{volume}{17}},
  \bibinfo{pages}{2343} (\bibinfo{year}{2007}).

\bibitem[{\citenamefont{Perra et~al.}(2009)\citenamefont{Perra, Zlati\'c,
  Chessa, Conti, Donato, and Caldarelli}}]{caldarelli}
\bibinfo{author}{\bibfnamefont{N.}~\bibnamefont{Perra}},
  \bibinfo{author}{\bibfnamefont{V.}~\bibnamefont{Zlati\'c}},
  \bibinfo{author}{\bibfnamefont{A.}~\bibnamefont{Chessa}},
  \bibinfo{author}{\bibfnamefont{C.}~\bibnamefont{Conti}},
  \bibinfo{author}{\bibfnamefont{D.}~\bibnamefont{Donato}}, \bibnamefont{and}
  \bibinfo{author}{\bibfnamefont{G.}~\bibnamefont{Caldarelli}},
  \bibinfo{journal}{Europhys. Lett.} \textbf{\bibinfo{volume}{88}},
  \bibinfo{pages}{48002} (\bibinfo{year}{2009}).

\bibitem[{\citenamefont{Bak et~al.}(2002)\citenamefont{Bak, Christensen, Danon,
  and Scanlon}}]{BakEarthquakes}
\bibinfo{author}{\bibfnamefont{P.}~\bibnamefont{Bak}},
  \bibinfo{author}{\bibfnamefont{K.}~\bibnamefont{Christensen}},
  \bibinfo{author}{\bibfnamefont{L.}~\bibnamefont{Danon}}, \bibnamefont{and}
  \bibinfo{author}{\bibfnamefont{T.}~\bibnamefont{Scanlon}},
  \bibinfo{journal}{Phys. Rev. Lett.} \textbf{\bibinfo{volume}{88}},
  \bibinfo{pages}{178501} (\bibinfo{year}{2002}).

\bibitem[{\citenamefont{Wheatland and Litvinenko}(2002)}]{solarFlares}
\bibinfo{author}{\bibfnamefont{M.~S.} \bibnamefont{Wheatland}}
  \bibnamefont{and} \bibinfo{author}{\bibfnamefont{Y.~E.}
  \bibnamefont{Litvinenko}}, \bibinfo{journal}{Sol. Phys.}
  \textbf{\bibinfo{volume}{211}}, \bibinfo{pages}{255} (\bibinfo{year}{2002}).

\bibitem[{\citenamefont{de~Arcangelis et~al.}(2006)\citenamefont{de~Arcangelis,
  C., Lippiello, and M.}}]{lucilla2}
\bibinfo{author}{\bibfnamefont{L.}~\bibnamefont{de~Arcangelis}},
  \bibinfo{author}{\bibfnamefont{C.}~\bibnamefont{Godano}},
  \bibinfo{author}{\bibfnamefont{E.}~\bibnamefont{Lippiello}},
  \bibnamefont{and} \bibinfo{author}{\bibfnamefont{M.}~\bibnamefont{Nicodemi}},
  \bibinfo{journal}{Phys. Rev. Lett.} \textbf{\bibinfo{volume}{96}},
  \bibinfo{pages}{051102} (\bibinfo{year}{2006}).

\bibitem[{\citenamefont{de~Arcangelis et~al.}(2010)\citenamefont{de~Arcangelis,
  C., Lippiello, and M.}}]{lucilla3}
\bibinfo{author}{\bibfnamefont{E.}~\bibnamefont{Lippiello}},
  \bibinfo{author}{\bibfnamefont{L.}~\bibnamefont{de~Arcangelis}},
  \bibnamefont{and} \bibinfo{author}{\bibfnamefont{C.}~\bibnamefont{Godano.}},
  \bibinfo{journal}{Astron.\& Astrophys.} \textbf{\bibinfo{volume}{488}},
  \bibinfo{pages}{L29} (\bibinfo{year}{2008}).

\bibitem[{\citenamefont{de~Arcangelis et~al.}(2010)\citenamefont{de~Arcangelis,
  C., Lippiello, and M.}}]{lucilla4}
\bibinfo{author}{\bibfnamefont{E.}~\bibnamefont{Lippiello}},
  \bibinfo{author}{\bibfnamefont{L.}~\bibnamefont{de~Arcangelis}},
  \bibnamefont{and} \bibinfo{author}{\bibfnamefont{C.}~\bibnamefont{Godano.}},
  \bibinfo{journal}{Astron.\& Astrophys.} \textbf{\bibinfo{volume}{511}},
  \bibinfo{pages}{L2} (\bibinfo{year}{2010}).

\bibitem[{\citenamefont{Corral et~al.}(2008)\citenamefont{Corral, Telesca, and
  Lasaponara}}]{forestFires}
\bibinfo{author}{\bibfnamefont{A.}~\bibnamefont{Corral}},
  \bibinfo{author}{\bibfnamefont{L.}~\bibnamefont{Telesca}}, \bibnamefont{and}
  \bibinfo{author}{\bibfnamefont{R.}~\bibnamefont{Lasaponara}},
  \bibinfo{journal}{Phys. Rev. E} \textbf{\bibinfo{volume}{77}},
  \bibinfo{pages}{016101} (\bibinfo{year}{2008}).

\bibitem[{\citenamefont{Varga}(2006)}]{inter-arr1}
\bibinfo{author}{\bibfnamefont{P.}~\bibnamefont{Varga}},
  \bibinfo{journal}{EUNICE 2005: Networks and Applications Towards a
  Ubiquitously Connected World, IFIP International Federation for Information
  Processing} \textbf{\bibinfo{volume}{196}}, \bibinfo{pages}{17}
  (\bibinfo{year}{2006}).

\bibitem[{\citenamefont{Zimmermann et~al.}(2005)\citenamefont{Zimmermann,
  Clark, Mohay, Pouget, and Dacier}}]{inter-arr2}
\bibinfo{author}{\bibfnamefont{J.}~\bibnamefont{Zimmermann}},
  \bibinfo{author}{\bibfnamefont{A.}~\bibnamefont{Clark}},
  \bibinfo{author}{\bibfnamefont{G.}~\bibnamefont{Mohay}},
  \bibinfo{author}{\bibfnamefont{F.}~\bibnamefont{Pouget}}, \bibnamefont{and}
  \bibinfo{author}{\bibfnamefont{M.}~\bibnamefont{Dacier}},
  \bibinfo{journal}{First International Workshop on Systematic Approaches to
  Digital Forensic Engineering} p.~\bibinfo{pages}{89} (\bibinfo{year}{2005}).

\bibitem[{\citenamefont{Pica~Ciamarra et~al.}(2008)\citenamefont{Pica~Ciamarra,
  Cognilio, and de~Arcangelis}}]{lucillaSpam}
\bibinfo{author}{\bibfnamefont{M.}~\bibnamefont{Pica~Ciamarra}},
  \bibinfo{author}{\bibfnamefont{A.}~\bibnamefont{Cognilio}}, \bibnamefont{and}
  \bibinfo{author}{\bibfnamefont{L.}~\bibnamefont{de~Arcangelis}},
  \bibinfo{journal}{Europhysics Letters} \textbf{\bibinfo{volume}{84}},
  \bibinfo{pages}{28004} (\bibinfo{year}{2008}).

\bibitem[{\citenamefont{Omori}(1894)}]{omoriLaw}
\bibinfo{author}{\bibfnamefont{F.}~\bibnamefont{Omori}}, \bibinfo{journal}{J.
  Coll. Sci. Imp. Univ. Tokyo} \textbf{\bibinfo{volume}{7}},
  \bibinfo{pages}{111} (\bibinfo{year}{1894}).

\bibitem[{\citenamefont{Molloy and B.}(1995)}]{configModel}
\bibinfo{author}{\bibfnamefont{M.}~\bibnamefont{Molloy}} \bibnamefont{and}
  \bibinfo{author}{\bibfnamefont{R.}~\bibnamefont{B.}},
  \bibinfo{journal}{Random Struct. Algorithms} \textbf{\bibinfo{volume}{6}},
  \bibinfo{pages}{161} (\bibinfo{year}{1995}).

\bibitem[{\citenamefont{Newman et~al.}(2001)\citenamefont{Newman, Strogatz, and
  D.J.}}]{newmanCM}
\bibinfo{author}{\bibfnamefont{M.~E.~J.} \bibnamefont{Newman}},
  \bibinfo{author}{\bibfnamefont{S.}~\bibnamefont{Strogatz}}, \bibnamefont{and}
  \bibinfo{author}{\bibfnamefont{W.}~\bibnamefont{D.J.}},
  \bibinfo{journal}{Phys. Rev. E} \textbf{\bibinfo{volume}{64}},
  \bibinfo{pages}{026118} (\bibinfo{year}{2001}).

\bibitem[{\citenamefont{Bollob\'as}(1985)}]{Bollobas}
\bibinfo{author}{\bibfnamefont{B.}~\bibnamefont{Bollob\'as}},
  \emph{\bibinfo{title}{Random Graphs}} (\bibinfo{publisher}{Academic Press,
  Orlando}, \bibinfo{year}{1985}).

\bibitem[{\citenamefont{Corral}(2004)}]{corralInterarrival}
\bibinfo{author}{\bibfnamefont{A.}~\bibnamefont{Corral}},
  \bibinfo{journal}{Phys. Rev. Lett.} \textbf{\bibinfo{volume}{92}},
  \bibinfo{pages}{108501} (\bibinfo{year}{2004}).

\bibitem[{\citenamefont{Kittas et~al.}(2008)\citenamefont{Kittas, Carmi,
  Havlin, and Argyrakis}}]{shlomoTrapping}
\bibinfo{author}{\bibfnamefont{A.}~\bibnamefont{Kittas}},
  \bibinfo{author}{\bibfnamefont{S.}~\bibnamefont{Carmi}},
  \bibinfo{author}{\bibfnamefont{S.}~\bibnamefont{Havlin}}, \bibnamefont{and}
  \bibinfo{author}{\bibfnamefont{P.}~\bibnamefont{Argyrakis}},
  \bibinfo{journal}{Europhysics Letters} \textbf{\bibinfo{volume}{84}},
  \bibinfo{pages}{40008} (\bibinfo{year}{2008}).

\end{thebibliography}
\end{document}